\documentclass[submission, Phys]{SciPost}

\hypersetup{
    colorlinks,
    linkcolor={red!50!black},
    citecolor={blue!50!black},
    urlcolor={blue!80!black}
}

\usepackage[bitstream-charter]{mathdesign}
\DeclareSymbolFont{usualmathcal}{OMS}{cmsy}{m}{n}
\DeclareSymbolFontAlphabet{\mathcal}{usualmathcal}

\begin{document}

\begin{center}{\Large\textbf{Quantum tricriticality of incommensurate phase induced by quantum strings in frustrated Ising magnetism\\
}}\end{center}

\begin{center}
Zheng Zhou\textsuperscript{1,2,3},
Changle Liu\textsuperscript{4,5,6},
Dong-Xu Liu\textsuperscript{6,7},
Zheng Yan\textsuperscript{8,9,2$\star$},
Yan Chen\textsuperscript{2,10$\dagger$},
Xue-Feng Zhang\textsuperscript{6,7$\ddagger$}
\end{center}

\begin{center}
{\bf 1} Perimeter Institute for Theoretical Physics, Waterloo, Ontario, Canada N2L 2Y5\\
{\bf 2} Department of Physics and State Key Laboratory of Surface Physics, Fudan University, Shanghai 200438, China\\
{\bf 3} Department of Physics and Astronomy, University of Waterloo, Waterloo, Ontario, Canada N2L 3G1\\
{\bf 4} School of Engineering, Dali University, Dali, Yunnan 671003, China\\
{\bf 5} Shenzhen Institute for Quantum Science and Technology and Department of Physics, Southern University of Science and Technology, Shenzhen 518055, China\\
{\bf 6} Department of Physics, and Center of Quantum Materials and Devices, Chongqing University, Chongqing, 401331, China\\
{\bf 7} Chongqing Key Laboratory for Strongly Coupled Physics, Chongqing University, Chongqing 401331, China\\
{\bf 8} Department of Physics and HKU-UCAS Joint Institute of Theoretical and Computational Physics, The University of Hong Kong, Pokfulam Road, Hong Kong, China\\
{\bf 9} Beihang Hangzhou Innovation Institute, Yuhang, Hangzhou 310023, China\\
{\bf 10} Collaborative Innovation Center of Advanced Microstructures, Nanjing 210093, China\\
$\star$ {\small \sf zhengyan@hku.hk},
$\dagger$ {\small \sf yanchen99@fudan.edu.cn},
$\ddagger$ {\small \sf zhangxf@cqu.edu.cn},
\end{center}

\begin{center}
\today
\end{center}

\section*{Abstract}
{\bf Incommensurability plays a critical role in many strongly correlated systems. In some cases, the origin of such exotic order can be theoretically understood in the framework of 1d line-like topological excitations known as ``quantum strings''. Here we study an extended transverse field Ising model on a triangular lattice. Using the large scale quantum Monte Carlo simulations, we find that the spatial anisotropy can stabilize an incommensurate phase out of the commensurate clock order. Our results for the structure factor and the string density exhibit a linear relationship between incommensurate ordering wave vector and
the density of quantum strings, which is reminiscent of hole density in under-doped cuprate superconductors.
When introducing the next-nearest-neighbor interaction, we observe a quantum tricritical point out of the incommensurate phase. After carefully analyzing the ground state energies within different string topological sectors, we conclude that this tricriticality is non-trivially caused by effective long-range inter-string interactions with two competing terms following different decaying behaviors.}

\vspace{10pt}
\noindent\rule{\textwidth}{1pt}
\tableofcontents\thispagestyle{fancy}
\noindent\rule{\textwidth}{1pt}
\vspace{10pt}

\section{Introduction}

Frustrated magnetism is the frontier of condensed matter physics and has been under active research in the recent decades. In frustrated magnets, competing interactions among local moments can give rise to exotic ground states and low-energy dynamics \cite{fru_1,fru_2,Co_1,Co_2,Co_3,Ba_1,Ba_2,Cs_1,Sg,Vb}. As one of the simplest models, the nearest-neighbor antiferromagnetic Ising model on various frustrated lattices exhibit extensive ground state degeneracy, hence has finite residue entropy \cite{TIM_R1, TIM_R2, TIM_R3}. When introducing quantum fluctuations such as transverse spin exchange interaction\cite{Co_1, Co_2, Co_3} and transverse field \cite{TIM_R1, TIM_R2, TIM_R3, TIM_Y1, TIM_Y2}, the extensive degeneracy will be lifted, and the system either enters an ordered phase with spontaneously broken symmetries, or a quantum disordered phase, depending on the microscopic details of the system.
However, in some cases quantum fluctuations are not enough to completely remove the extensive degeneracy. As an example, systems with spatial exchange anisotropy \cite{Sl_2, Ic_1} or further-neighbor interactions \cite{rk_zhou} can stabilize \textit{incommensurte} ordering that still exhibits sub-extensive degeneracy.

The incommensurability (or high-order commensurability) is usually reminiscent of the famous high-$T_c$ cuprate superconductors \cite{SC_1, SC_6}. When a Mott-insulator with antiferromagnetic order is doped with holes, the string-type topological excitations will be constructed and divide the whole system into many domains, so that an incommensurate order is formed \cite{Zaanen_89, irrelcite2, SC_2, SC_3, SC_4}.
On the other hand, string excitations have been observed in cold atom optical lattices that simulates the Hubbard model \cite{SCO_1, SCO_2}.
Moreover, an incommensurate supersolid phase has been theoretically predicted for the repulsive hard-core bosons trapped in the anisotropic triangular optical lattice \cite{Zhang_1}. However, the incommensurate phase induced by the string-type topological defects is still not fully understood, especially the related quantum phase transition which is more experimentally feasible \cite{SC_4, SCO_3, SC_5}.

\begin{figure}[btp!]
    \centering
    \includegraphics[width=.7\linewidth]{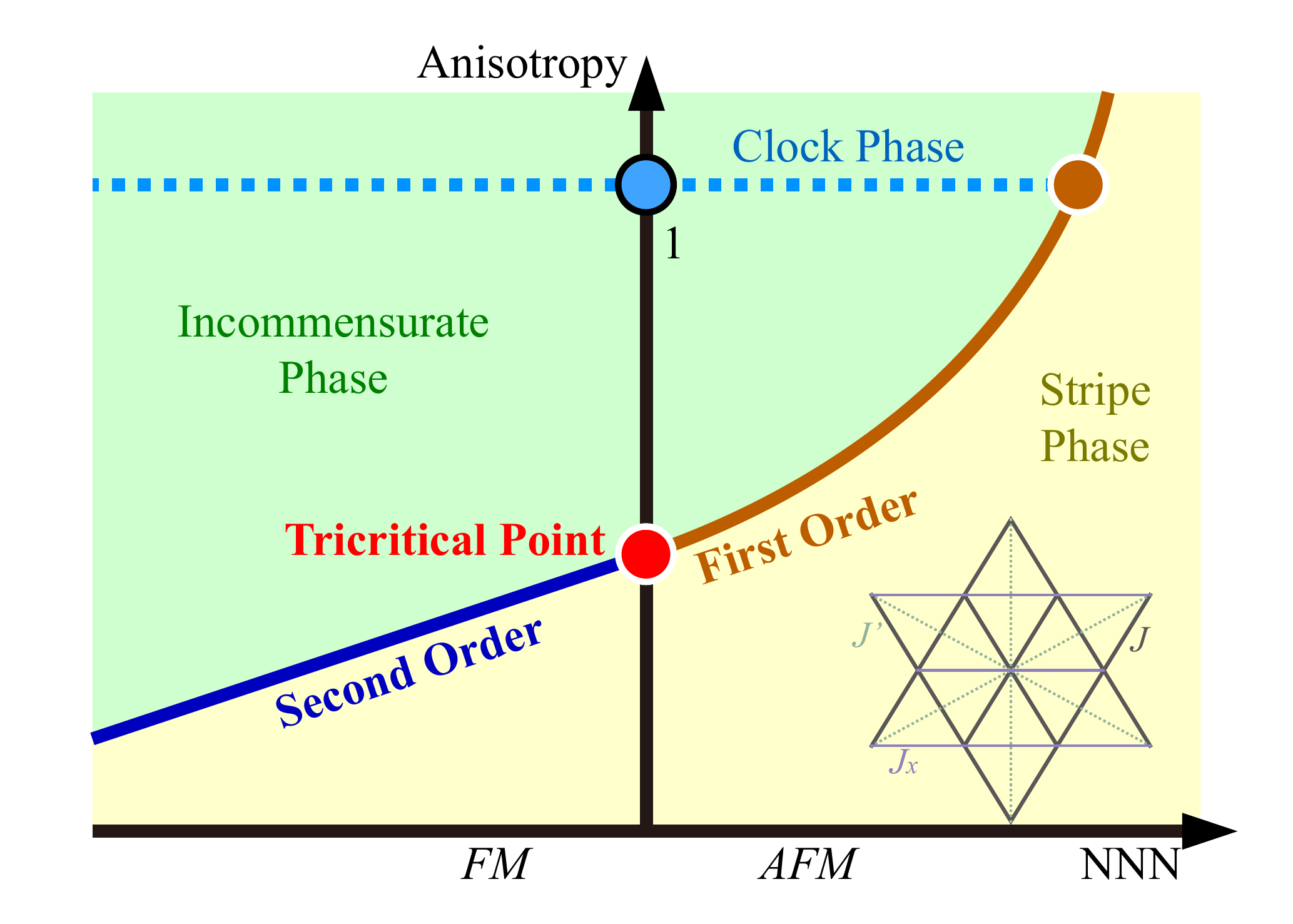}
    \caption{Schematic phase diagram of the extended TFIM versus next-nearest-neighbor interaction $J'/J$ and spatial anisotropy $J_x/J$. Inset: an illustration of different interactions.}
    \label{fig1}
\end{figure}

In this paper, we study the transverse field Ising model (TFIM) in an anisotropic triangular lattice, which is highly related to several strongly correlated systems such as rare-earth frustrated magnets \cite{Li_2020,shen_2019,wei_2020}, antiferroelectric materials \cite{Chai_2016}, trapped ions \cite{Trap_2012} and Rydberg atom arrays \cite{Scholl2021,Ebadi2021}. As shown in Fig. \ref{fig1}, in the presence of a spatial anisotropy, the clock phase is extended to be part of a quantum incommensurate phase, which can be understood with the quantum topological defect -- string-type domain wall excitations.
When a next-nearest neighbor (NNN) interaction is introduced, a quantum tricritical point emerges on the incommensurate--stripe transition line. Non-trivially, such tricritical behavior is found due to the changing of effective interactions between the quantum strings.

The rest of the paper is organized as follows. In Section \ref{Sec_QS}, we introduce the anisotropic TFIM,
and demonstrate that the low-energy physics of this model can be well formulated in terms of quantum strings.
In Section \ref{Sec_QMC}, we present our numerical results for the quantum strings. In Section \ref{Sec_tri}, we show that introducing NNN interaction can give rise to a quantum tricriticality. We also demonstrate such non-trivial critical behavior is caused by an effective long-range inter-string interaction with two competing terms. In Section \ref{Sec_clock}, we consider the isotropic TFIM with finite NNN interaction, which the relevant with recent experiments.

\section{Quantum strings in the anisotropic quantum Ising model}
\label{Sec_QS}

In this paper we investigate the anisotropic TFIM with nearest-neighbor anisotropic interactions. The Hamiltonian takes the following form
\begin{equation}
    H=J_x\sum_{\langle ij\rangle_x}S_i^zS_j^z+J\sum_{\langle ij\rangle_\wedge}S_i^zS_j^z-h\sum_{i}S_i^x,
    \label{Eq:ham}
\end{equation}
where $h$ is the transverse field, and $\langle ij\rangle_x$ and $\langle ij\rangle_\wedge$ denote the horizontal and diagonal inter-chain bonds, respectively.
The TFIM in the isotropic limit $J=J_x$ has been well understood with renormalization group \cite{TIM_RG1,TIM_RG2}, effective field theory \cite{TIM_Y1,TIM_Y2} and quantum Monte Carlo simulations \cite{TIM_R1,TIM_R2,TIM_R3}.
On the other hand, the incommensurability is found in the XXZ model in the anisotropic triangular lattice \cite{Zhang_1}. Therefore, it is intuitive to believe that the incommensurability also exits in our anisotropic TFIM Eq. \eqref{Eq:ham}.

\begin{figure}[btp!]
    \centering
    \includegraphics[width=.7\linewidth]{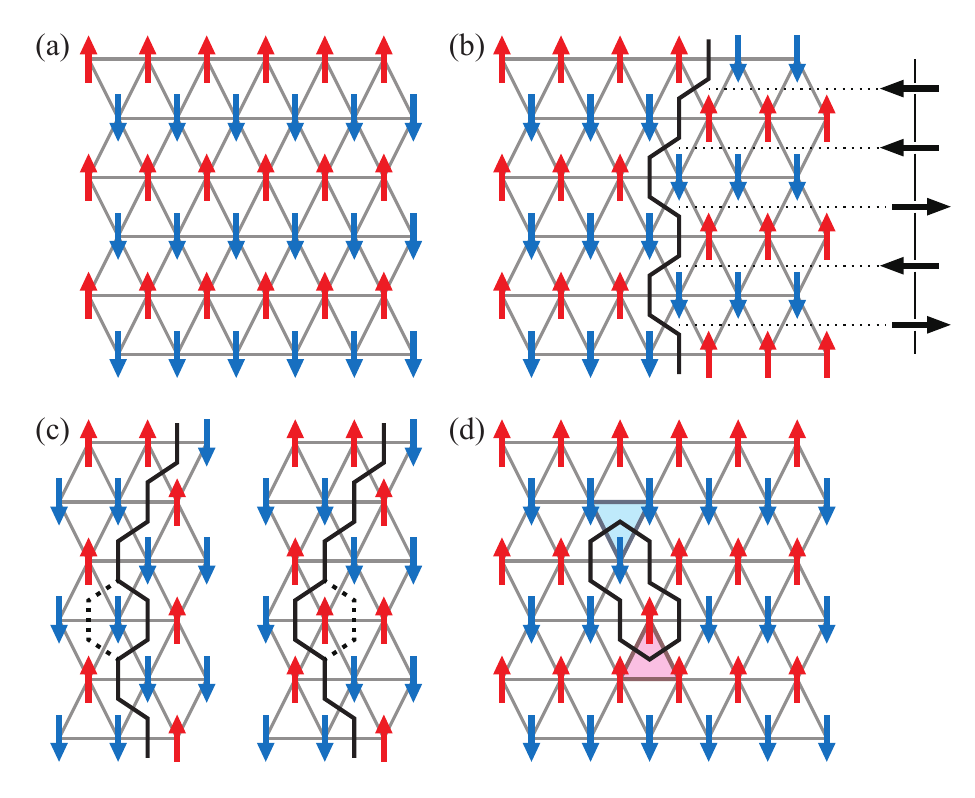}
    \caption{(a) The spin configuration in stripe phase. (b) The spin configuration with one quantum string which can be mapped to spin-$1/2$ chain. (c) The deformation of the string by flipping spin. (d) The spinon excitations which break the triangle-rule. }
    \label{fig2}
\end{figure}

We consider the Ising limit of $h=0$ firstly. To minimize the ground state energy, spins within each triangle need to satisfy the triangle-rule, \textit{i.e.}, one of the spins must be opposite to the other two. The triangle rule here is similar to the ice-rule in pyrochlore spin ice systems \cite{SI}. In the classical anisotropic case $J_x<J$ and $h=0$, all the bonds connecting parallel spins are aligned along the $x$-direction so that a two-fold stripe order is formed, as shown in Fig. \ref{fig2}(a). However, at the isotropic limit $J_x=J$, the bonds connecting parallel spins have no preferred directions, hence there will be extensive ground state degeneracy associated with finite residue entropy \cite{Pauling} $S/N=0.323k_\mathrm{B}$ \cite{TIM_R1,TIM_R2,TIM_R3}.
As demonstrated in Fig.\ref{fig2}(b),
we can express spin configurations in terms of domains of the stripe states, if we take the two-fold stripe states as reference states. The reason we choose the stripe state as reference is that it is the ground state of the anisotropic model. Moreover, it exhibits no low-energy dynamics, \textit{i.e.}, any local operations above the stripe states violate the triangle rule and do not survive at low energies, hence the low-energy dynamics only exist at the domain walls. To avoid creation of triangle-rule-breaking defects (also known as `spinon' topological defects), each bisector within the string can only choose left-going or right-going directions. With this setup, the low-energy spin configurations can be formulated in terms of the directed domain walls or strings that penetrates through the $y$ direction of the lattice. An example of such mapping is shown in Fig. \ref{fig2}(b). Meanwhile, the illustration also shows that the internal dynamics of a single string can be exactly mapped to an effective spin-$1/2$ chain if we map the left-going (right-going) bisector to effective spin-up (down) \cite{TIM_Y1, TIM_Y2, Zhang_2, Wan_1, Wan_2}. However, such string is still `classical' and have an energy gap $\Delta=L_y(J-J_x)/2$.

The transverse field $h$ introduces quantum fluctuations to the system, the strings will fluctuate as is illustrated in Fig. \ref{fig2}(c).
With the first-order perturbation theory, we find that at low energies, the transverse field leads to nearest-neighbor XY interactions within the effective spin-$1/2$ chain,
and the strings become `quantum'. On the other hand, from Fig. \ref{fig2}(d) we find that the transverse field term can also generate spinon topological defects that break the triangle-rule.
However, such process is suppressed at low energies since the creation of spinon excitations cost a large energy of $\sim J$ \cite{TIM_Y1, TIM_Y2,Zheng_1}.

In the low-temperature and small transverse field region $T, |h|\ll J$, the low-energy Hilbert space of the Hamiltonian Eq. \eqref{Eq:ham} is approximately restricted to the spinon-free sector that can be well formulated in terms of directed strings. As discussed above, a single string can be mapped to an effective spin-$1/2$ XY chain, a model exactly solvable with Jordan-Wigner transformation. The energy of a single quantum string turns out to be
\begin{eqnarray}
\nonumber
E_\mathrm{QS}&=&\Delta+E_\mathrm{XY}\\
&=&L_y(J-J_x)/2-L_yh/\pi.
\end{eqnarray}
Then, $E_\mathrm{QS}$ can be taken as the effective chemical potential of the strings. The string excitation will be energy preferred compared with the stripe phase when $h/\pi>(J-J_x)/2$. However, the number of strings can not explosively increase when $E_{\mathrm{QS}}<0$, because of the effective interactions between them.

\begin{figure}[btp!]
    \centering
    \includegraphics[width=.7\linewidth]{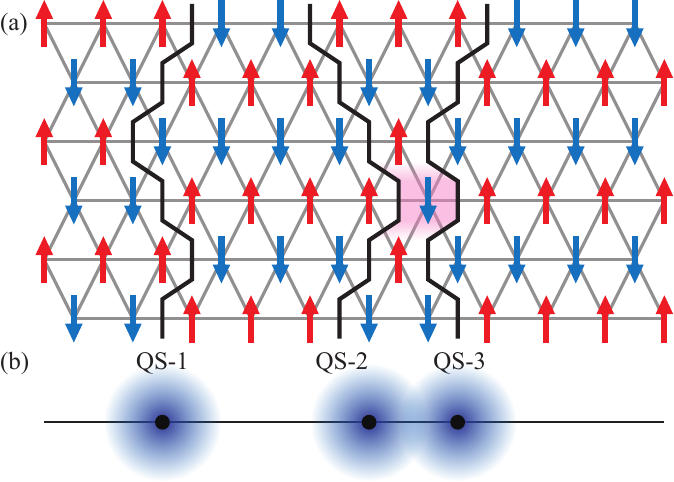}
    \caption{(a) A spin configuration that is mapped to three strings. At low energies, the strings can not cross with each other (the spin shaded with red cannot flip due to triangle-rule), so that there exists effective interactions between strings. (b) The configuration in upper panel can be taken as three hard-core bosons with long-range effective interactions in one dimension.}
    \label{fig3}
\end{figure}

When multiple strings are presented in the system, their interactions become significant to the quantum excitations. The non-crossing condition \cite{TIM_Y1, TIM_Y2} which states that different strings cannot intersect first imposes a kinematic hard-core constraint to strings. The vibration of strings~[Fig.~\ref{fig2}(c)] then turns this kinematic hard-core constraint to a dynamic repulsion. Specifically, when two strings come close to each other (e.g., the QS-2 and QS-3 in Fig.~\ref{fig3}a), their vibration will be constrained, reducing the energy gain from internal vibration. This loss of vibration energy can be equivalently regarded as the repulsive interaction between them. The dynamic repulsion can be described by a power-law long-range potential.
With a mean-field treatment, the average energy of quantum string can be written as
\begin{equation}
    \bar{E}=L_y\left(\frac{J-J_x}{2}-\frac{h}{\pi}\right)+V(\bar{r}).
\end{equation}
where $V(r)$ defines the effective repulsive energy between two strings separated by a distance $r$, and $\bar{r}$ is the mean distance between nearby strings. However, it is hard to obtain the explicit expression of function $V(r)$  analytically.
Numerical simulations must be performed to determine the form of $V(r)$.

\section{Numerical results for the quantum strings}
\label{Sec_QMC}
We carry out a quantum Monte Carlo (QMC) calculation using stochastic series expansion (SSE) algorithm \cite{sandvik_1997,sandvik_1998,sandvik_2003,avella_2013}. We adopt a rectangular periodic boundary condition in both directions. In the simulation, $10^5$ and $2.9\times 10^6$ Monte Carlo steps are used for equilibration and measurements, respectively. Since the updates between different topological sectors (with different number of strings) is difficult in QMC at ultra-low temperatures, we equilibrate the configurations into different topological sectors and compare their energies to find the global ground state. The unit of energy scale is set to be $J=1$. To distinguish different quantum phases, we choose two observables -- structure factors and density of quantum strings.
The spin structure factor is defined as the Fourier transform of the spin-spin correlator in the Ising component:
\begin{equation}
S(\mathbf{q})=\frac{1}{N}\sum_{ij}\langle S_i^zS_j^z\rangle\exp [i\mathbf{q}\cdot(\mathbf{r}_i-\mathbf{r}_j)],
\end{equation}
which characterizes different magnetic orders.
And the average density of quantum strings is defined as
\begin{equation}
\rho_\mathrm{QS}=\frac{1}{N}\sum_{\langle ij\rangle_x}\frac{1-4\langle S_i^zS_j^z\rangle}{2}
\end{equation}
which effectively counts the number of horizontal bonds connecting opposite spins.

\begin{figure}[btp!]
    \centering
    \includegraphics[width=1\linewidth]{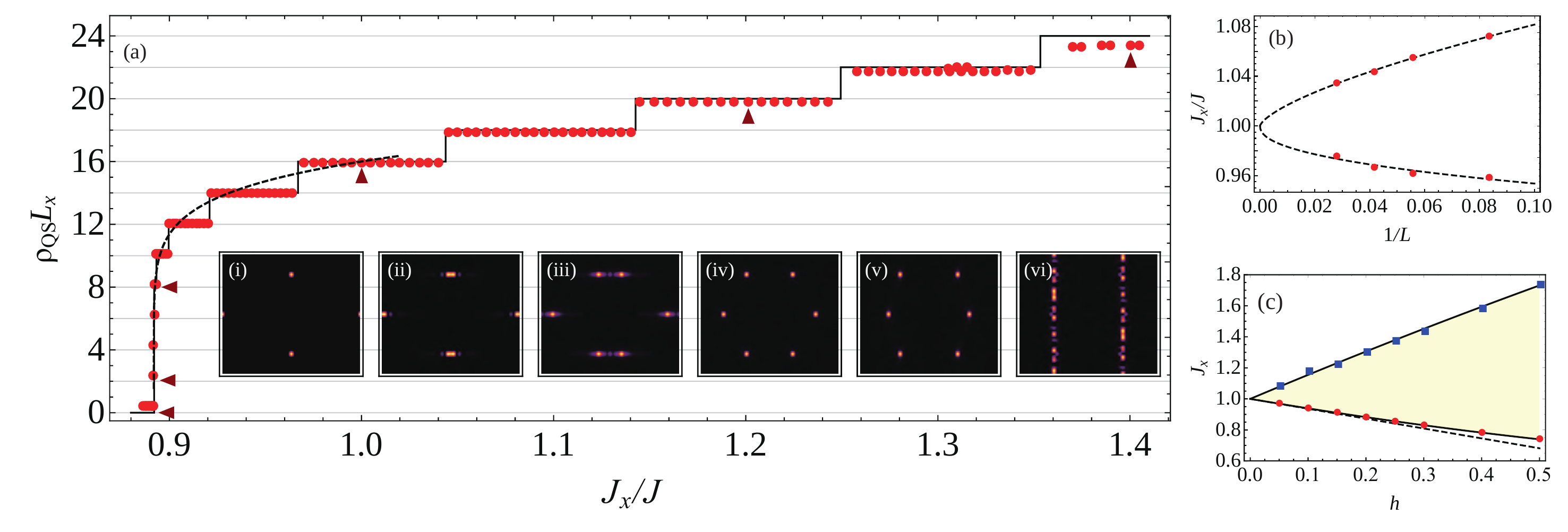}
    \caption{(a) The relation between quantum string density $\rho_\mathrm{QS}$ and anisotropy $J_x/J$, measured on a $24\times 24$ lattice at $h=0.2$ and $\beta=24$. The dashed line gives the theoretical result in the thermodynamic limit. The grey gridlines show the plateaux of quantization corresponding to even numbers of strings. Inset: The structure factor at the parameter region with 0, 2, 8, 16, 20, and 24 strings. The panel (i), (iv), and (vi) corresponds to stripe, clock phase, and decoupled chains, respectively. The dashed white hexagon denotes the first Brillouin zone. (b) The finite-size scaling result for the upper and lower boundary of the $\rho_\mathrm{QS}=2/3$ plateau. (c) The phase diagram on the $h$--$J_x$ plane. The yellowish region denotes the incommensurate phase. The upper and lower critical line separates it with the decoupled chain phase and stripe phase, respectively.}
    \label{fig4}
\end{figure}

In the weak $x$-coupling limit $J_x\ll J$, as shown in Fig. \ref{fig4}(a), the quantum string density $\rho_{\mathrm{QS}}$ is approximately zero, indicating the formation of the ferromagnetic order in the $x$-direction. Meanwhile, the peaks of structure factor $S(\mathbf{q})$ at wave vectors $\mathbf{q}=(\pm 2\pi,0)$ and $(0,\pm 2\pi/\sqrt{3})$ in inset (i) of Fig. \ref{fig4}(a) confirms the existence of stripe phase. The ordering momentum is commensurate. The spin alignment alternates each row in the real space configuration~[Fig.~\ref{fig2}(a)]. This phase extends to $J_x\to-\infty$ and persists at the classical limit $h=0$. When increasing the coupling $J_x/J$, a jump of $\rho_{\mathrm{QS}}$ is observed at $J_{x,\mathrm{c}}=0.892$ which is close to the theoretical estimation $J-2h/\pi=0.873$ where the single string energy $E_{\textrm{QS}}$ becomes zero.

Across $J_{x,c}$ a series of plateaux can be clearly found, which hints strong topological characteristics. The values $\rho_\mathrm{QS}L_x$
indicate the total number of strings $N_\mathrm{QS}$ of the global ground state. Meanwhile, we notice that only even number of strings exist, and this is due to the periodical boundary conditions that we adopt in the simulation. As the TFIM on a triangular lattice can be mapped to the quantum dimer model \cite{TIM_R1, TIM_R2, TIM_R3, Yan}, it can be shown that spin configurations with different string density correspond to dimer configurations with different winding numbers.

The magnetic ordering of multiple-string states can be examined by the spin structure factors, as shown in the insets of Fig. \ref{fig4}. The clear and sharp magnetic Bragg peaks indicate well-established magnetic ordering. Meanwhile, when increasing the $\rho_{\mathrm{QS}}$, the centers of peaks move from M points along $x$-direction to K points at $\rho_\mathrm{QS}=2/3$, and continue to move towards $(\pm\pi,0)$ as $\rho_\mathrm{QS}>2/3$. Such behaviors can be understood once we consider the interactions between the strings. If the interaction $V(r)$ is repulsive and decays with distance, the quantum strings will tend to distribute with equal distance in $x$-direction on average. Then, the emergence of quantum strings can introduce additional modulation with wave vectors to be \textit{incommensurate} or called high-order commensurate $\mathbf{q}=(\pm(2-N_\mathrm{QS}/L_x),0)\pi$ and $(\pm N_\mathrm{QS}/L_x,\pm 2/\sqrt{3})\pi$, which are verified by $S(\mathbf{q})$ with high accuracy in our QMC simulations.

Looking closely at the plateaux, we can find their widths increase with $\rho_{\mathrm{QS}}$. It indicates that the repulsive interaction $V(r)$ decays faster than the linear type. To get the explicit form of the interactions, we consider the total energy
\begin{equation}
E(\rho_\mathrm{QS})=N\rho_\mathrm{QS}\left(\frac{J-J_x}{2}-\frac{h}{\pi}+f(\rho_\mathrm{QS})\right)\label{Er}
\end{equation}
where $f(\rho_\mathrm{QS})=V(\bar{r})/L_y$ and $\bar{r}=L_x/N_{\mathrm{QS}}=1/\rho_{\mathrm{QS}}$. Previous work \cite{Zhang_1} points out that this interaction should be algebraic $f(\rho_\mathrm{QS})\propto\rho_\mathrm{QS}^\alpha$ in the thermodynamic limit. Thus, by extremizing the total energy ${\partial E/}{\partial\rho_\mathrm{QS}}=0$, we get the QS density $\rho_\mathrm{QS}(J_x)\propto(J_x-J_{x,c})^{1/\alpha}$. Then, with help of Eq. \ref{Er}, we obtain the exponent $\alpha\approx7.5(1)$ by fitting the jumping points between different plateaux. Hereafter, because the clock phase at isotropic point $J_x=J$ can also be taken as special commensurate state of quantum strings with $\rho_{\mathrm{QS}}=2/3$, we can get the explicit form of QS density without fitting the proportional coefficient
\begin{equation}
\rho_\mathrm{QS}(J_x)=\frac{2}{3}\left(\frac{J_x-J_{x,c}}{J-J_{x,c}}\right)^{1/\alpha}, \label{rdm}
\end{equation}
which matches well with the finite size numerical results in Fig. \ref{fig4}.

The possible number of plateaux increases while increasing the system size. Meanwhile, their widths shrink. As demonstrated in finite-size scaling (Fig. \ref{fig4}(b)) of the plateau corresponding to the clock phase, the width of each plateaux scales to zero the string density $\rho_\mathrm{QS}$ should follow the continuous function~(\ref{rdm}) in the thermodynamic limit. In the meantime, the continuous changes of $\rho_\mathrm{QS}$ from $0$ to $1$ and position of peaks $\mathbf{q}$ from $(\pm 2\pi,0)$ to $(\pm\pi,0)$, show a clear mark of continuous phase transition. Besides the clock phase, the state with $\mathbf{q}=(\pm\pi, q_y)$ at $J_x>J$ is also special. The spins form a spin-density wave in $x$-direction on each horizontal chain, but the inter-chain interactions are so weak that all chains are decoupled in $y$ direction. Thus, the maximum of structure factor locate at $q_x=\pm\pi$ and arbitrary $q_y$ which marks absence of order in $y$-direction (inset (vi) of Fig.~\ref{fig4}).

At last, we also take different $h$ values and calculate phase diagram in the $h$--$J_x/J$ plane (Fig. \ref{fig4}c). The region of the incommensurate phase takes the shape of a sector centered at $h=0, J_x=1$ point, and expands broader as $h$ increases. For relatively small $h$, the critical lines are linear and agrees with the theoretical value. For larger $h$, the critical line gradually deviates, since higher-order perturbation is not negligible.

\begin{figure}[btp!]
	\centering
	\includegraphics[width=\linewidth]{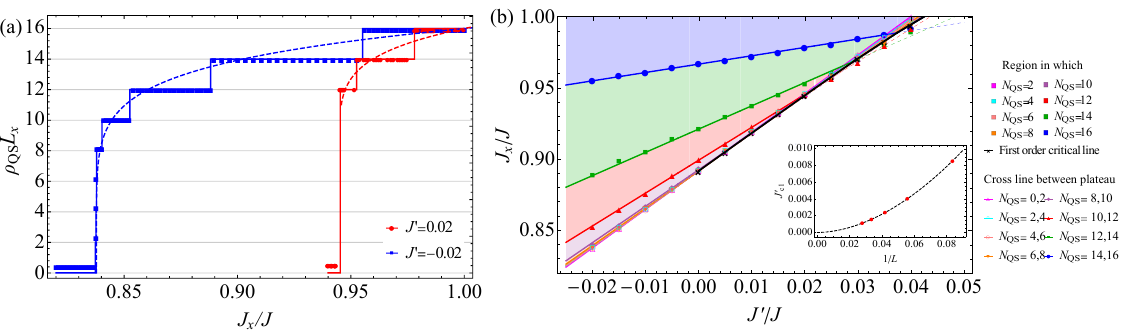}
	\caption{(a) Quantum string density $\rho_\mathrm{QS}$ versus anisotropy $J_x/J$ relation with NNN interaction at $L=\beta=24$ and $h/J=0.2$. The dashed line gives the theoretical results in thermodynamic limit. (b) The numerical finite size phase diagram in NNN interaction--anisotropy plane measured at $L=\beta=24$ and $h/J=0.2$. The filled regions shows parameter regions with different number of strings and the colored lines shows the jumping lines between them. The black line shows the first order transition line. Inset: The expected behavior of the disappearing point of the first plateau at finite size towards thermodynamic limit given by $E(\rho_\mathrm{QS}=0)=E(\rho_\mathrm{QS}=2/L)=E(\rho_\mathrm{QS}=4/L)$.}
	\label{fig6}
\end{figure}

\section{Tricriticality}
\label{Sec_tri}
In the frustrated magnetic materials \cite{Li_2020,shen_2019}, trapped ions \cite{Trap_2012} and the Rydberg atom arrays  \cite{Scholl2021,Ebadi2021}, the influence of long-range interactions can not be neglected. In the Ising limit, such interaction can break the extensive degeneracy and result in the stripe phase. However, different from continuous transition through the incommensurate phase in the anisotropic case, the recent QMC simulation \cite{wei_2020} points out the direct first-order phase transition from the stripe phase to clock phase. Thus, an intrinsic quantum tricriticality is expected and related to the interplay between long-range interaction and spatial anisotropy. To figure it out, we introduce a next-nearest-neighbor (NNN) interaction with magnitude $J'$ into the Hamiltonian:
\begin{equation}
    H=J_x\sum_{\langle ij\rangle_x}S_i^zS_j^z+J\sum_{\langle ij\rangle_\wedge}S_i^zS_j^z+J'\sum_{\langle\langle ij\rangle\rangle}S_i^zS_j^z-h\sum_{i}S_i^x \label{H_nnn}
\end{equation}
where $\langle\langle ij\rangle\rangle$ denotes all the NNN sites.

The NNN interaction can be tuned from ferromagnetic to antiferromagnetic, and it can strongly change the type of the quantum phase transition reflected by the relation between string density and anisotropy. As shown in Fig. \ref{fig6}(a), all the plateaux can be detected at the presence of a ferromagnetic NNN interaction, just like no NNN interaction case in Fig. \ref{fig4}(a).
However, in contrast to such clear evidence of the continuous phase transition, we can find a notable jump in the antiferromagnetic side. To be specific, for $L=\beta=24$ and $J'=h/10=0.02$ (red dots in Fig \ref{fig6}(a)), as $J_x$ decrease from 1, the number of strings $N_\mathrm{QS}$ changes from 16 to 12, then directly jumps to 0 and skips all the plateaux ranging from $N_\mathrm{QS}=2$ to $10$. Thus, in the thermodynamic limit, we expect the existence of a first-order transition point $J_{x,c}$ where string density jump from a critical value $\rho_c$ to zero for antiferromagnetic NNN interactions.

The different types of incommensurate--stripe phase transition indicates that a tricritical point should locate around $J'/J\approx0$. Usually, the tricritical point is related to the order parameter field with multi-components, such as coplanar phase \cite{tri_xxz1,tri_xxz2}, or binary Bose mixtures \cite{two-comp}. However, here the tricriticality has its topological feature. It divides the parameter region where topological sectors of strings are complete or not. Such a feature is similar to the incommensurate phase in high-$T_c$ superconductivity, where the first-order phase transition is probably due to the NNN interactions between hole domain walls \cite{SC_6}.

To locate the tricritical point, we map out the phase diagram in the $J_x$--$J'$ plane, as is shown in Fig. \ref{fig6}(b). For finite size systems, the incommensurate phase is divided into regions with quantized numbers of strings, separated by jumping lines where the energy of $N_\mathrm{QS}=k$ and $k+2$ is equal. The transition line is determined in such a way that the energy of the stripe phase equals the lowest energy of the incommensurate phase among all the quantized plateaux. For negative $J'$, all the jumping lines are well determined. While $J'$ increases to be positive, the jumping lines gradually cross the transition line. Accompanying crossing, the plateaux will disappear one after another. Thus, the first plateau $N_\mathrm{QS}=2$ will disappear at finite size tricritical point which can be obtain by solve the equation $E(\rho_\mathrm{QS}=0;J',J_x)=E(\rho_\mathrm{QS}=2/L_x;J',J_x)=$ $E(\rho_\mathrm{QS}=4/L_x;J',J_x)$. Moving towards the thermodynamic limit, this point is found to gradually approach $J'=0$ in Fig. \ref{fig6}(b) inset. Then, the question turns to how to understand such tricriticality through the effective paradigm of quantum strings.

Taken into account the energy intra- and inter-strings, our effective model yields that the energy of strings can be written as $E(\rho_\mathrm{QS})\propto\rho_\mathrm{QS}(A+B\rho_\mathrm{QS}^\alpha)$, which is always concave and only have one extremum. However, to create two competing phases apart requires at least two extrema, so the ansatz of interactions between strings needs to be adjusted at the presence of the NNN interaction.

\begin{figure}[btp!]
    \centering
    \includegraphics[width=\linewidth]{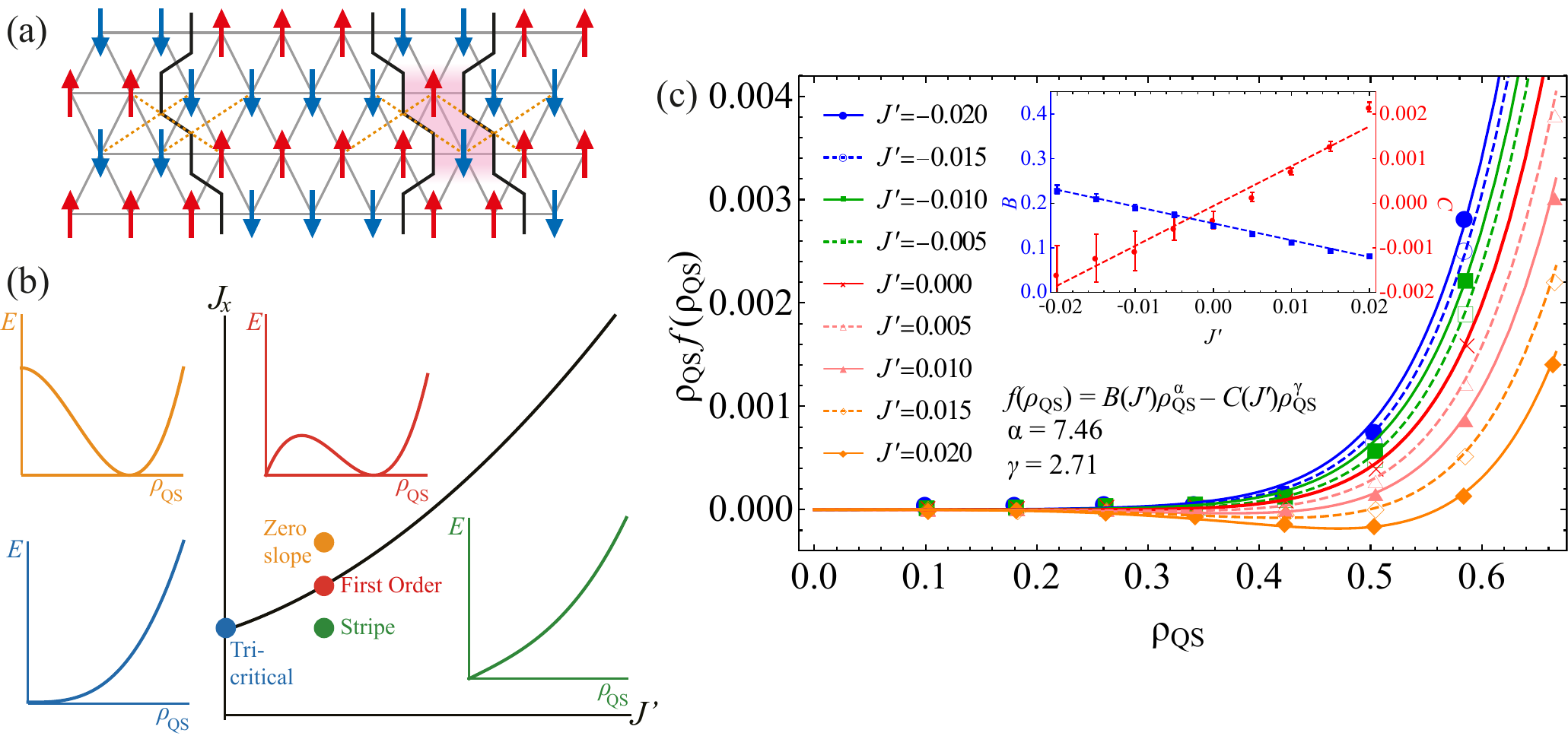}
    \caption{(a) The spin configuration with three strings. The orange lines show the intra- and inter-string energy induced by NNN interactions. (b) An illustration of the energy versus string density curve in vicinity of the tricritical point. (c) The numerical result of QS interaction versus QS density relation for different NNN interaction $J'$. The dashed and solid lines show the result from the fitting. Inset: The interaction coefficient $B$ and $C$ versus NNN interaction $J'$.}
    \label{fig7}
\end{figure}

In fact, at the presence of $J'$, there is an additional mechanism of string interaction:
apart from the repulsion from the hinder of motion when strings are nearby denoted $V_h(r)$, the second, denoted $V_{J'}(r)$, comes from the fact that the insertion of single string produces energy cost $3J'/2$ per string length, while two adjacent strings cost energy $2J'$ per string length, which is different than two individual strings~[Fig.~\ref{fig7}(a)]; therefore we can write $V(r)=V_h(r)+V_{J'}(r)$. Originally, these two mechanisms all act in short-range; the vibration of strings then turns these short-range interactions into long-range ones. The exact determination of the form of $V_{J'}(r)$ and $V_h(r)$ is a difficult question. In a former work~\cite{Zhang_1}, the repulsion $V_h(r)$ from motion hinder has been determined to follow a power-law behaviour in the incommensurate phase a similar model (hardcore Bose-Hubbard model) by fitting the jumping points of the plateaux. As an extension, it is most natural and simplest to assume that $V_{J'}$ also follow a power law. We therefore write down the ansatz $V(r)=B(J')/r^\alpha-C(J')/r^\gamma$, and a straightforward term $-C(J')\rho_\mathrm{QS}^{\gamma}$ be added into the energy of string
\begin{equation}
E(\rho_\mathrm{QS})=N\rho_\mathrm{QS}(A+B(J')\rho_\mathrm{QS}^\alpha-C(J')\rho_\mathrm{QS}^\gamma).
\label{E_nnn}
\end{equation}

Because the energy above should return to no NNN interaction case when setting $J'=0$, we can get $C(0)=0$ and $B(0)\ne0$. Keeping the leading order, Eq. \ref{E_nnn} should be rewritten as
\begin{equation}
E(\rho_\mathrm{QS})\approx N\rho_\mathrm{QS}(A+B_0\rho_\mathrm{QS}^\alpha-C_1J'\rho_\mathrm{QS}^\gamma).
\end{equation}
If $C_1$ is positive and $\alpha>\gamma$, the appearance of tricriticality can be well explained, as demonstrated in Fig. \ref{fig7}(b). When $J'$ is small and positive, the two competing interactions result in an inflection point in the $E$--$\rho_\mathrm{QS}$ curve. At a small $J_x$, the stripe phase has the lowest energy. Then, when increasing $J_x$, due to the last term of $E(\rho_{\mathrm{QS}})$, another energy minimum appears for the incommensurate phase with certain non-zero quantum string density. The energy gap to the stripe phase will be closed at transition point $J_{x,c}$. The jumping of quantum string density $\rho_{\mathrm{QS}}$ indicates phase transition is the first order. When $J'<0$, the last term takes the same effect as the second term, and the incommensurate--stripe phase transition remains second order. We also note that for a more accurate description, $B(J')$ should be approximated as $B(J')\approx B_0+B_1J'$ to take into account the fact that the NNN coupling modifies the vibration of string\footnote{In particular, the effective model of a single string will be modified from an XY chain to an XXZ chain.}, but this modification does not change the scenario qualitatively and can be therefore considered as a higher-order term.

In order to numerically determine the interaction coefficients in
$f(\rho_{\mathrm{QS}})=B(J')\rho_{\mathrm{QS}}^\alpha-C(J')$ $\rho_{\mathrm{QS}}^\gamma$, we calculated the energy $E(\rho_{\mathrm{QS}})$ versus string density
for different $J'$. In finite size system, the linear term can be obtained by finding $E_l=E(\rho_{\mathrm{QS}}=0)= E(\rho_{\mathrm{QS}}=2/L_x)$ (which equals the critical point when $J'<0$). Then, after substracting the linear term from $E(\rho_{\mathrm{QS}})$, we can get the finite size value of $\rho_{\mathrm{QS}}f(\rho_{\mathrm{QS}})$. As shown in Fig. \ref{fig7}(c), $\rho_{\mathrm{QS}}f(\rho_{\mathrm{QS}})$ fits well with the exponents $\alpha=7.5(1)$ and $\gamma=2.7(3)$, and $\alpha$ larger than $\gamma$ coincides with our prediction. Meanwhile, from the inset of Fig. \ref{fig7}(c), we can find the leading order approximation of the coefficients $B$ and $C$ are good enough. Thus, the numerical results strongly support our theoretical analysis: \textit{the tricriticality of the incommensurate phase is caused by the competing of effective long-range inter-string interactions with different power exponents.}

\section{Clock phase}
\label{Sec_clock}
\begin{figure}[tbh!]
	\centering
	\includegraphics[width=.7\linewidth]{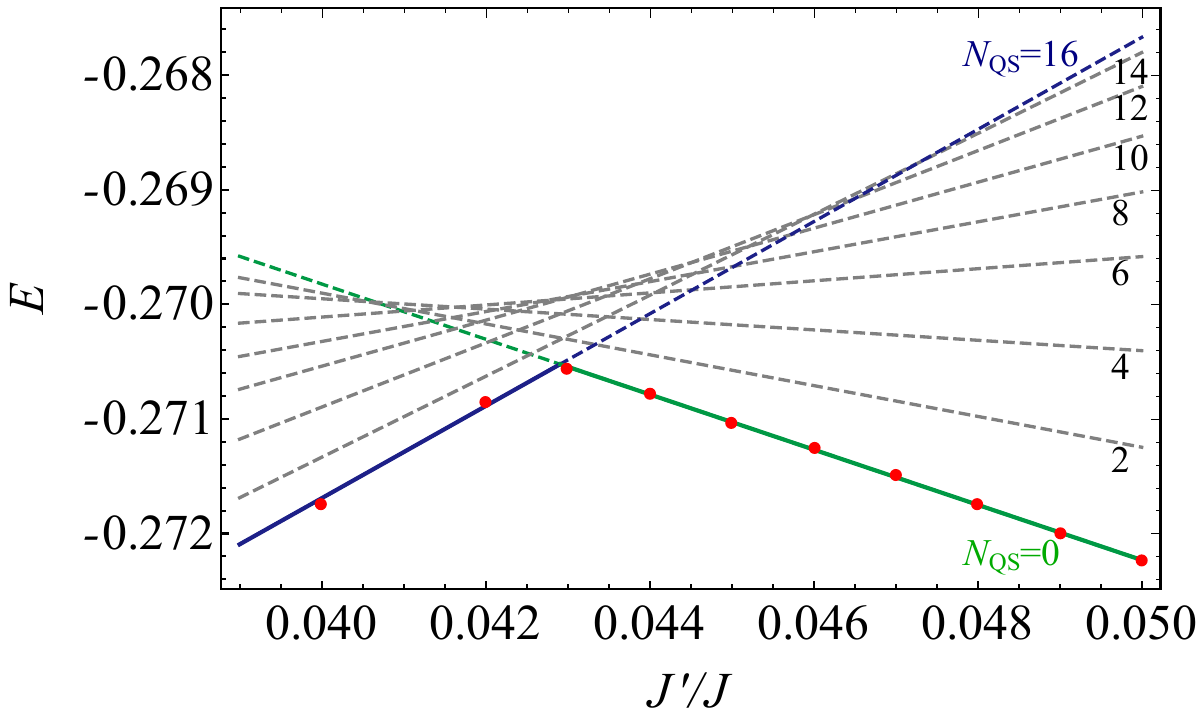}
	\caption{Ground state energies within each topological sectors as a function of the NNN interaction $J'$ at $h/J=0.2$. The solid lines and the red dots denote the global minimum energies, and the gray dashed lines denote the energy of excited states with different numbers of strings.}
	\label{fig9}
\end{figure}

Recently, there have been some experimental and numerical studies on the isotropic TFIM with NNN interaction \cite{Li_2020,shen_2019,wei_2020}. Without the longitudinal magnetic field, the ground state is found to be the antiferromagnetic clock phase. Meanwhile, the numerical simulation discovers an excitation mode, which is identified as `roton'. Based on the effective field theory \cite{TIM_Y1, TIM_Y2} and the renormalization group \cite{TIM_RG1, TIM_RG2}, the thermal phase transition from the clock phase to the Kosterlitz-Thouless phase should be of the Berezinskii-Kosterlitz-Thouless (BKT) type. On the other hand, the Kosterlitz-Thouless phase will undergo an another BKT phase transition to the paramagnetic phase as increasing temperatures, and this transition is driven by proliferation of vortices which is dual to spinons that break the triangle-rule. Here we turn to the isotropic case $J_x=J$ of Hamiltonian Eq. \eqref{H_nnn}.

For small $h/J$, the effective theory of strings can still work well \cite{TIM_Y1, TIM_Y2}. In Fig. \ref{fig9}, we find a first-order transition between the clock phase (with string density $\rho_{\mathrm{QS}}=2/3$) and the stripe phase (with string density $\rho_{\mathrm{QS}}=0$). All the incommensurate states with $0<\rho_{\mathrm{QS}}<2/3$ are higher-lying excited states that are irrelevant at the phase transition.
Recently,
the excitation spectra within different phases have been simulated with the stochastic analytic continuation technique \cite{Zheng_1}. Quantum dynamics associated with the quantum strings have been well analyzed, and it strongly supports the quantum strings as a valid description of the low-energy physics of the TFIM in the triangular lattice. We also note that at the first-order transition point, the energies of different topological sectors are quite close with energy diference per site $\Delta E\sim 10^{-3}$. This feature is reminiscent of the elusive Rokhsar-Kivelson point~\cite{rk_debut} in quantum dimer models. In fact, it has been suggested that by adding a small third-nearest-neighbor Ising coupling term in Eq. \eqref{H_nnn}, this first-order transition can be driven into a Rokhsar-Kivelson deconfined quantum critical point~\cite{rk_zhou}.
Therefore, we expect that at the finite temperature window $T\gtrsim\Delta E$, quantum critical behaviors of Rokhsar-Kivelson point can be observed above this first-order transition point.

\section{Conclusion and Outlook}

To summarize, we study the extended TFIM on a triangular lattice with spatial exchange anisotropy and NNN interactions. We find that an incommensurate order emerges in presence of the spatial anisotropy, which can be well understood in terms of quantum strings.
The phase transition between the incommensurate phase and the stripe phase is found to be continuous.
Then, we check the influence of the NNN interaction $J'$ and find the phase transition remains continuous when $J'<0$ and becomes first order when $J'>0$. There exists a quantum tricritical point separating these two regions approximately at zero NNN interaction point $J'=0$. To figure out the mechanism, we propose a new ansatz of the effective inter-string interaction with form $\frac{B}{r^{\alpha}}-\frac{C}{r^{\gamma}}$, where $r$ is the distance between two nearby strings. By fitting with the numerical results, we obtain the power exponents are $\alpha=7.5(1)$ and $\gamma=2.7(3)$, and the leading order approximation is good enough for explaining such exotic quantum tricriticality. Last, we revisit the isotropic case with NNN interaction, and find the clock state can also be understood via quantum strings.

Our work provides insightful predictions of the real system. For frustrated magnetism, the incommensurate phase should be found when introducing the spatial anisotropy, such as adding pressure. On the other hand, the spatial anisotropy and also the NNN interaction can be flexible tuned in the programmable Rydberg atom arrays \cite{Scholl2021,Ebadi2021}, so the tricritical point could be verified. Besides, the first-order phase transition resulting from the NNN interaction may shed light on a similar effect on the high-$T_c$ cuprate superconductors.

\section*{Acknowledgements}

We wish to thank Yang Qi, Jun Zhao, Zi-Yang Meng, Yin-Chen He, and Frank Pollmann for the fruitful discussions. This work was supported by the National Key Research and Development Program of China (Grants No. 2017YFA0304204 and No. 2016YFA0300504), the National Natural Science Foundation of China (Grants No. 11625416 and No. 11474064), and the Shanghai Municipal Government (Grants No. 19XD1400700 and No. 19JC1412702). X.-F. Z. acknowledges funding from the National Science Foundation of China under Grants  No. 11874094 and No.12147102, Fundamental Research Funds for the Central Universities Grant No. 2021CDJZYJH-003. Z. Z. acknowledges support from the CURE (H.-C. Chin and T.-D. Lee Chinese Undergraduate Research Endowment) (19925) and National University Student Innovation Program (201910246148). C. L. is supported by the National Natural Science Foundation of China (11925402), Guangdong province (2016ZT06D348, 2020KCXTD001), the National Key R \& D Program (2016YFA0301700), Shenzhen High-level Special Fund (G02206304, G02206404), and the Science, Technology and Innovation Commission of Shenzhen Municipality (ZDSYS20170303165926217, JCYJ20170412152620376, KYTDPT20181011104202253), and Center for Computational Science and Engineering of SUSTech. The authors acknowledge Beijing PARATERA Tech CO.,Ltd. (\url{https://www.paratera.com/}) for providing HPC resources that have contributed to the research results reported within this paper.


\begin{thebibliography}{10}
    \providecommand{\url}[1]{\texttt{#1}}
    \providecommand{\urlprefix}{URL }
    \expandafter\ifx\csname urlstyle\endcsname\relax
      \providecommand{\doi}[1]{doi:\discretionary{}{}{}#1}\else
      \providecommand{\doi}{doi:\discretionary{}{}{}\begingroup
      \urlstyle{rm}\Url}\fi
    \providecommand{\eprint}[2][]{\url{#2}}

    \bibitem{fru_1}
    R.~Moessner and A.~P. Ramirez,
    \newblock \emph{Geometrical frustration},
    \newblock Phys. Today \textbf{59}(2), 24 (2006),
    \newblock \doi{10.1063/1.2186278}.

    \bibitem{fru_2}
    M.~Vojta,
    \newblock \emph{Frustration and quantum criticality},
    \newblock Rep. Prog. Phys. \textbf{81}(6), 064501 (2018),
    \newblock \doi{10.1088/1361-6633/aab6be}.

    \bibitem{Co_1}
    D.~Sellmann, X.-F. Zhang and S.~Eggert,
    \newblock \emph{Phase diagram of the antiferromagnetic {XXZ} model on the
      triangular lattice},
    \newblock Phys. Rev. B \textbf{91}, 081104 (2015),
    \newblock \doi{10.1103/PhysRevB.91.081104}.

    \bibitem{Co_2}
    D.~Yamamoto, G.~Marmorini and I.~Danshita,
    \newblock \emph{Quantum phase diagram of the triangular-lattice {XXZ} model in
      a magnetic field},
    \newblock Phys. Rev. Lett. \textbf{112}, 127203 (2014),
    \newblock \doi{10.1103/PhysRevLett.112.127203}.

    \bibitem{Co_3}
    O.~A. Starykh, W.~Jin and A.~V. Chubukov,
    \newblock \emph{Phases of a triangular-lattice antiferromagnet near
      saturation},
    \newblock Phys. Rev. Lett. \textbf{113}, 087204 (2014),
    \newblock \doi{10.1103/PhysRevLett.113.087204}.

    \bibitem{Ba_1}
    T.~Susuki, N.~Kurita, T.~Tanaka, H.~Nojiri, A.~Matsuo, K.~Kindo and H.~Tanaka,
    \newblock \emph{Magnetization process and collective excitations in the
      ${S}=1/2$ triangular-lattice heisenberg antiferromagnet
      $\mathrm{{B}a_3{C}o{S}b_2{O}_9}$},
    \newblock Phys. Rev. Lett. \textbf{110}, 267201 (2013),
    \newblock \doi{10.1103/PhysRevLett.110.267201}.

    \bibitem{Ba_2}
    Y.~Kamiya, L.~Ge, T.~Hong, Y.~Qiu, D.~Quintero-Castro, Z.~Lu, H.~Cao,
      M.~Matsuda, E.~Choi, C.~Batista \emph{et~al.},
    \newblock \emph{The nature of spin excitations in the one-third magnetization
      plateau phase of $\mathrm{{B}a_3{C}o{S}b_2{O}_9}$},
    \newblock Nat. Commun. \textbf{9}, 2666 (2018),
    \newblock \doi{10.1038/s41467-018-04914-1}.

    \bibitem{Cs_1}
    N.~A. Fortune, S.~T. Hannahs, Y.~Yoshida, T.~E. Sherline, T.~Ono, H.~Tanaka and
      Y.~Takano,
    \newblock \emph{Cascade of magnetic-field-induced quantum phase transitions in
      a spin-$\frac{1}{2}$ triangular-lattice antiferromagnet},
    \newblock Phys. Rev. Lett. \textbf{102}, 257201 (2009),
    \newblock \doi{10.1103/PhysRevLett.102.257201}.

    \bibitem{Sg}
    Z.~Ma, J.~Wang, Z.-Y. Dong, J.~Zhang, S.~Li, S.-H. Zheng, Y.~Yu, W.~Wang,
      L.~Che, K.~Ran, S.~Bao, Z.~Cai \emph{et~al.},
    \newblock \emph{Spin-glass ground state in a triangular-lattice compound
      $\mathrm{{Y}b{Z}n{G}a{O}_4}$},
    \newblock Phys. Rev. Lett. \textbf{120}, 087201 (2018),
    \newblock \doi{10.1103/PhysRevLett.120.087201}.

    \bibitem{Vb}
    J.~P. Sheckelton, J.~R. Neilson, D.~G. Soltan and T.~M. McQueen,
    \newblock \emph{Possible valence-bond condensation in the frustrated cluster
      magnet $\mathrm{{L}i{Z}n_2{M}o_3{O}_8}$},
    \newblock Nat. Mat. \textbf{11}(6), 493 (2012),
    \newblock \doi{10.1038/nmat3329}.

    \bibitem{TIM_R1}
    R.~Moessner, S.~L. Sondhi and P.~Chandra,
    \newblock \emph{Two-dimensional periodic frustrated {I}sing models in a
      transverse field},
    \newblock Phys. Rev. Lett. \textbf{84}, 4457 (2000),
    \newblock \doi{10.1103/PhysRevLett.84.4457}.

    \bibitem{TIM_R2}
    R.~Moessner and S.~L. Sondhi,
    \newblock \emph{{I}sing models of quantum frustration},
    \newblock Phys. Rev. B \textbf{63}, 224401 (2001),
    \newblock \doi{10.1103/PhysRevB.63.224401}.

    \bibitem{TIM_R3}
    S.~V. Isakov and R.~Moessner,
    \newblock \emph{Interplay of quantum and thermal fluctuations in a frustrated
      magnet},
    \newblock Phys. Rev. B \textbf{68}, 104409 (2003),
    \newblock \doi{10.1103/PhysRevB.68.104409}.

    \bibitem{TIM_Y1}
    Y.~Jiang and T.~Emig,
    \newblock \emph{String picture for a model of frustrated quantum magnets and
      dimers},
    \newblock Phys. Rev. Lett. \textbf{94}, 110604 (2005),
    \newblock \doi{10.1103/PhysRevLett.94.110604}.

    \bibitem{TIM_Y2}
    Y.~Jiang and T.~Emig,
    \newblock \emph{Ordering of geometrically frustrated classical and quantum
      triangular {I}sing magnets},
    \newblock Phys. Rev. B \textbf{73}, 104452 (2006),
    \newblock \doi{10.1103/PhysRevB.73.104452}.

    \bibitem{Sl_2}
    Y.~Yoshida, H.~Ito, M.~Maesato, Y.~Shimizu, H.~Hayama, T.~Hiramatsu,
      Y.~Nakamura, H.~Kishida, T.~Koretsune, C.~Hotta \emph{et~al.},
    \newblock \emph{Spin-disordered quantum phases in a quasi-one-dimensional
      triangular lattice},
    \newblock Nat. Phys. \textbf{11}(8), 679 (2015),
    \newblock \doi{10.1038/nphys3359}.

    \bibitem{Ic_1}
    A.~Weichselbaum and S.~R. White,
    \newblock \emph{Incommensurate correlations in the anisotropic triangular
      heisenberg lattice},
    \newblock Phys. Rev. B \textbf{84}, 245130 (2011),
    \newblock \doi{10.1103/PhysRevB.84.245130}.

    \bibitem{rk_zhou}
    Z.~Zhou, Z.~Yan, C.~Liu, Y.~Chen and X.-F. Zhang,
    \newblock \emph{Emergent {R}okhsar-{K}ivelson point in realistic quantum
      {I}sing models},
    \newblock \doi{10.48550/ARXIV.2106.05518} (2021).

    \bibitem{SC_1}
    E.~Dagotto,
    \newblock \emph{Correlated electrons in high-temperature superconductors},
    \newblock Rev. Mod. Phys. \textbf{66}, 763 (1994),
    \newblock \doi{10.1103/RevModPhys.66.763}.

    \bibitem{SC_6}
    R.~J.~Birgeneau, C.~Stock, J.~M.~Tranquada and K.~Yamada,
    \newblock \emph{Magnetic neutron scattering in hole-doped cuprate
      superconductors},
    \newblock J. Phys. Soc. Jpn. \textbf{75}(11), 111003 (2006),
    \newblock \doi{10.1143/JPSJ.75.111003}.

    \bibitem{Zaanen_89}
    J.~Zaanen and O.~Gunnarsson,
    \newblock \emph{Charged magnetic domain lines and the magnetism of
      high-${T}_{c}$ oxides},
    \newblock Phys. Rev. B \textbf{40}, 7391 (1989),
    \newblock \doi{10.1103/PhysRevB.40.7391}.

    \bibitem{irrelcite2}
    M.~Kato, K.~Machida, H.~Nakanishi and M.~Fujita,
    \newblock \emph{Soliton lattice modulation of incommensurate spin density wave
      in two dimensional {H}ubbard model-a mean field study},
    \newblock J. Phys. Soc. Jpn. \textbf{59}(3), 1047 (1990),
    \newblock \doi{10.1143/JPSJ.59.1047}.

    \bibitem{SC_2}
    J.~Tranquada, B.~Sternlieb, J.~Axe, Y.~Nakamura and S.~Uchida,
    \newblock \emph{Evidence for stripe correlations of spins and holes in copper
      oxide superconductors},
    \newblock Nature \textbf{375}(6532), 561 (1995),
    \newblock \doi{10.1038/375561a0}.

    \bibitem{SC_3}
    B.-X. Zheng, C.-M. Chung, P.~Corboz, G.~Ehlers, M.-P. Qin, R.~M. Noack, H.~Shi,
      S.~R. White, S.~Zhang and G.~K.-L. Chan,
    \newblock \emph{Stripe order in the underdoped region of the two-dimensional
      {H}ubbard model},
    \newblock Science \textbf{358}(6367), 1155 (2017),
    \newblock \doi{10.1126/science.aam7127}.

    \bibitem{SC_4}
    Y.~Yu, L.~Ma, P.~Cai, R.~Zhong, C.~Ye, J.~Shen, G.~D. Gu, X.~H. Chen and
      Y.~Zhang,
    \newblock \emph{High-temperature superconductivity in monolayer
      $\mathrm{{B}i_2{S}r_2{C}a{C}u_2{O}}_{8+\delta}$},
    \newblock Nature \textbf{575}(7781), 156 (2019),
    \newblock \doi{10.1038/s41586-019-1718-x}.

    \bibitem{SCO_1}
    A.~Mazurenko, C.~S. Chiu, G.~Ji, M.~F. Parsons, M.~Kan{\'a}sz-Nagy, R.~Schmidt,
      F.~Grusdt, E.~Demler, D.~Greif and M.~Greiner,
    \newblock \emph{A cold-atom {F}ermi--{H}ubbard antiferromagnet},
    \newblock Nature \textbf{545}(7655), 462 (2017),
    \newblock \doi{10.1038/nature22362}.

    \bibitem{SCO_2}
    C.~S. Chiu, G.~Ji, A.~Bohrdt, M.~Xu, M.~Knap, E.~Demler, F.~Grusdt, M.~Greiner
      and D.~Greif,
    \newblock \emph{String patterns in the doped {H}ubbard model},
    \newblock Science \textbf{365}(6450), 251 (2019),
    \newblock \doi{10.1126/science.aav3587}.

    \bibitem{Zhang_1}
    X.-F. Zhang, S.~Hu, A.~Pelster and S.~Eggert,
    \newblock \emph{Quantum domain walls induce incommensurate supersolid phase on
      the anisotropic triangular lattice},
    \newblock Phys. Rev. Lett. \textbf{117}, 193201 (2016),
    \newblock \doi{10.1103/PhysRevLett.117.193201}.

    \bibitem{SCO_3}
    P.~T. Brown, E.~Guardado-Sanchez, B.~M. Spar, E.~W. Huang, T.~P. Devereaux and
      W.~S. Bakr,
    \newblock \emph{Angle-resolved photoemission spectroscopy of a
      {F}ermi--{H}ubbard system},
    \newblock Nat. Phys. \textbf{16}(1), 26 (2020),
    \newblock \doi{10.1038/s41567-019-0696-0}.

    \bibitem{SC_5}
    A.~Damascelli, Z.~Hussain and Z.-X. Shen,
    \newblock \emph{Angle-resolved photoemission studies of the cuprate
      superconductors},
    \newblock Rev. Mod. Phys. \textbf{75}, 473 (2003),
    \newblock \doi{10.1103/RevModPhys.75.473}.

    \bibitem{Li_2020}
    Y.~Li, S.~Bachus, H.~Deng, W.~Schmidt, H.~Thoma, V.~Hutanu, Y.~Tokiwa, A.~A.
      Tsirlin and P.~Gegenwart,
    \newblock \emph{Partial up-up-down order with the continuously distributed
      order parameter in the triangular antiferromagnet
      $\mathrm{{T}m{M}g{G}a{O}_4}$},
    \newblock Phys. Rev. X \textbf{10}, 011007 (2020),
    \newblock \doi{10.1103/PhysRevX.10.011007}.

    \bibitem{shen_2019}
    Y.~Shen, C.~Liu, Y.~Qin, S.~Shen, Y.-D. Li, R.~Bewley, A.~Schneidewind, G.~Chen
      and J.~Zhao,
    \newblock \emph{Intertwined dipolar and multipolar order in the
      triangular-lattice magnet $\mathrm{{T}m{M}g{G}a{O}_4}$},
    \newblock Nat. Commun. \textbf{10}(1), 1 (2019),
    \newblock \doi{10.1038/s41467-019-12410-3}.

    \bibitem{wei_2020}
    H.~Li, Y.~Da~Liao, B.-B. Chen, X.-T. Zeng, X.-L. Sheng, Y.~Qi, Z.~Y. Meng and
      W.~Li,
    \newblock \emph{{K}osterlitz-{T}houless melting of magnetic order in the
      triangular quantum {I}sing material $\mathrm{{T}m{M}g{G}a{O}_4}$},
    \newblock Nat. Commun. \textbf{11}(1), 1 (2020),
    \newblock \doi{10.1038/s41467-020-14907-8}.

    \bibitem{Chai_2016}
    S.-P. Shen, J.-C. Wu, J.-D. Song, X.-F. Sun, Y.-F. Yang, Y.-S. Chai, D.-S.
      Shang, S.-G. Wang, J.~F. Scott and Y.~Sun,
    \newblock \emph{Quantum electric-dipole liquid on a triangular lattice},
    \newblock Nat. Commun. \textbf{7}(1), 1 (2016),
    \newblock \doi{10.1038/ncomms10569}.

    \bibitem{Trap_2012}
    J.~W. Britton, B.~C. Sawyer, A.~C. Keith, C.-C.~J. Wang, J.~K. Freericks,
      H.~Uys, M.~J. Biercuk and J.~J. Bollinger,
    \newblock \emph{Engineered two-dimensional {I}sing interactions in a
      trapped-ion quantum simulator with hundreds of spins},
    \newblock Nature \textbf{484}(7395), 489 (2012),
    \newblock \doi{10.1038/nature10981}.

    \bibitem{Scholl2021}
    P.~{Scholl}, M.~{Schuler}, H.~J. {Williams}, A.~A. {Eberharter}, D.~{Barredo},
      K.-N. {Schymik}, V.~{Lienhard}, L.-P. {Henry}, T.~C. {Lang}, T.~{Lahaye},
      A.~M. {L{\"a}uchli} and A.~{Browaeys},
    \newblock \emph{{Quantum simulation of 2D antiferromagnets with hundreds of
      Rydberg atoms}},
    \newblock Nature \textbf{595}(7866), 233 (2021),
    \newblock \doi{10.1038/s41586-021-03585-1},
    \newblock \eprint{2012.12268}.

    \bibitem{Ebadi2021}
    S.~{Ebadi}, T.~T. {Wang}, H.~{Levine}, A.~{Keesling}, G.~{Semeghini},
      A.~{Omran}, D.~{Bluvstein}, R.~{Samajdar}, H.~{Pichler}, W.~W. {Ho},
      S.~{Choi}, S.~{Sachdev} \emph{et~al.},
    \newblock \emph{{Quantum phases of matter on a 256-atom programmable quantum
      simulator}},
    \newblock Nature \textbf{595}(7866), 227 (2021),
    \newblock \doi{10.1038/s41586-021-03582-4},
    \newblock \eprint{2012.12281}.

    \bibitem{TIM_RG1}
    D.~Blankschtein, M.~Ma, A.~N. Berker, G.~S. Grest and C.~M. Soukoulis,
    \newblock \emph{Orderings of a stacked frustrated triangular system in three
      dimensions},
    \newblock Phys. Rev. B \textbf{29}, 5250 (1984),
    \newblock \doi{10.1103/PhysRevB.29.5250}.

    \bibitem{TIM_RG2}
    J.~V. Jos\'e, L.~P. Kadanoff, S.~Kirkpatrick and D.~R. Nelson,
    \newblock \emph{Renormalization, vortices, and symmetry-breaking perturbations
      in the two-dimensional planar model},
    \newblock Phys. Rev. B \textbf{16}, 1217 (1977),
    \newblock \doi{10.1103/PhysRevB.16.1217}.

    \bibitem{SI}
    D.~Pomaranski, L.~Yaraskavitch, S.~Meng, K.~Ross, H.~Noad, H.~Dabkowska,
      B.~Gaulin and J.~Kycia,
    \newblock \emph{Absence of pauling’s residual entropy in thermally
      equilibrated $\mathrm{{D}y_2{T}i_2{O}_7}$},
    \newblock Nature Physics \textbf{9}(6), 353 (2013),
    \newblock \doi{10.1038/nphys2591}.

    \bibitem{Pauling}
    L.~Pauling,
    \newblock \emph{The structure and entropy of ice and of other crystals with
      some randomness of atomic arrangement},
    \newblock J. Am. Chem. Soc. \textbf{57}(12), 2680 (1935),
    \newblock \doi{10.1021/ja01315a102}.

    \bibitem{Zhang_2}
    X.-F. Zhang and S.~Eggert,
    \newblock \emph{Chiral edge states and fractional charge separation in a system
      of interacting bosons on a kagome lattice},
    \newblock Phys. Rev. Lett. \textbf{111}, 147201 (2013),
    \newblock \doi{10.1103/PhysRevLett.111.147201}.

    \bibitem{Wan_1}
    Y.~Wan and O.~Tchernyshyov,
    \newblock \emph{Quantum strings in quantum spin ice},
    \newblock Phys. Rev. Lett. \textbf{108}, 247210 (2012),
    \newblock \doi{10.1103/PhysRevLett.108.247210}.

    \bibitem{Wan_2}
    Y.~Wan, J.~Carrasquilla and R.~G. Melko,
    \newblock \emph{Spinon walk in quantum spin ice},
    \newblock Phys. Rev. Lett. \textbf{116}, 167202 (2016),
    \newblock \doi{10.1103/PhysRevLett.116.167202}.

    \bibitem{Zheng_1}
    Z.~Zhou, C.~Liu, Z.~Yan, Y.~Chen and X.-F. Zhang,
    \newblock \emph{Quantum dynamics of topological strings in a frustrated {I}sing
      antiferromagnet},
    \newblock npj Quantum Mater. \textbf{7}, 60 (2022),
    \newblock \doi{10.1038/s41535-022-00465-3}.

    \bibitem{sandvik_1997}
    A.~W. Sandvik,
    \newblock \emph{Finite-size scaling of the ground-state parameters of the
      two-dimensional heisenberg model},
    \newblock Phys. Rev. B \textbf{56}, 11678 (1997),
    \newblock \doi{10.1103/PhysRevB.56.11678}.

    \bibitem{sandvik_1998}
    A.~W. Sandvik,
    \newblock \emph{Stochastic method for analytic continuation of quantum {M}onte
      {C}arlo data},
    \newblock Phys. Rev. B \textbf{57}, 10287 (1998),
    \newblock \doi{10.1103/PhysRevB.57.10287}.

    \bibitem{sandvik_2003}
    A.~W. Sandvik,
    \newblock \emph{Stochastic series expansion method for quantum {I}sing models
      with arbitrary interactions},
    \newblock Phys. Rev. E \textbf{68}, 056701 (2003),
    \newblock \doi{10.1103/PhysRevE.68.056701}.

    \bibitem{avella_2013}
    A.~Avella and F.~Mancini,
    \newblock \emph{Strongly correlated systems: numerical methods}, vol. 176,
    \newblock Springer Science \& Business Media (2013).

    \bibitem{Yan}
    Z.~Yan, Y.~Wu, C.~Liu, O.~F. Sylju\aa{}sen, J.~Lou and Y.~Chen,
    \newblock \emph{Sweeping cluster algorithm for quantum spin systems with strong
      geometric restrictions},
    \newblock Phys. Rev. B \textbf{99}, 165135 (2019),
    \newblock \doi{10.1103/PhysRevB.99.165135}.

    \bibitem{tri_xxz1}
    D.~Sellmann, X.-F. Zhang and S.~Eggert,
    \newblock \emph{Phase diagram of the antiferromagnetic {XXZ} model on the
      triangular lattice},
    \newblock Phys. Rev. B \textbf{91}, 081104 (2015),
    \newblock \doi{10.1103/PhysRevB.91.081104}.

    \bibitem{tri_xxz2}
    D.~Yamamoto, G.~Marmorini and I.~Danshita,
    \newblock \emph{Quantum phase diagram of the triangular-lattice {XXZ} model in
      a magnetic field},
    \newblock Phys. Rev. Lett. \textbf{112}, 127203 (2014),
    \newblock \doi{10.1103/PhysRevLett.112.127203}.

    \bibitem{two-comp}
    Y.~Kato, D.~Yamamoto and I.~Danshita,
    \newblock \emph{Quantum tricriticality at the superfluid-insulator transition
      of binary bose mixtures},
    \newblock Phys. Rev. Lett. \textbf{112}, 055301 (2014),
    \newblock \doi{10.1103/PhysRevLett.112.055301}.

    \bibitem{rk_debut}
    D.~S. Rokhsar and S.~A. Kivelson,
    \newblock \emph{Superconductivity and the quantum hard-core dimer gas},
    \newblock Phys. Rev. Lett. \textbf{61}, 2376 (1988),
    \newblock \doi{10.1103/PhysRevLett.61.2376}.

    \end{thebibliography}
\end{document}